\documentclass[preprint,
 amsmath,amssymb,
 aps,
]{revtex4-2}

\usepackage{graphicx}% Include figure files
\usepackage{dcolumn}% Align table columns on decimal point
\usepackage{bm}% bold math
\usepackage{subcaption}
\usepackage{xcolor}
\begin{document}

% \preprint{IOP}%APS/123-QED

\title{Effects of adaptive acceleration response of birds on collective behaviors}% Force line breaks with \\
\author{Narina Jung}
\email{njung@kias.re.kr}
\affiliation{ School of Computational Sciences, Korea Institute for Advanced Study (KIAS), Seoul 02455, South Korea}

\author{Byung Mook Weon}%
\affiliation{Research Center for Advanced Materials Technology, Sungkyunkwan University, Suwon 16419, South Korea, and \\ Soft Matter Physics Laboratory\\School of Advanced Materials Science and Engineering\\ Sungkyunkwan University, Suwon 16419, South Korea}

\author{Pilwon Kim}
\email{pwkim@unist.ac.kr} 
\affiliation{Department of Mathematical Sciences\\ Ulsan National Institute of Science and Technology (UNIST), Ulsan 44919, South Korea}

% \collaboration{CLEO Collaboration}%\noaffiliation

\date{\today}% It is always \today, today,

\begin{abstract}
Collective dynamics of many interacting particles have been widely studied because of a wealth of their behavioral patterns quite different from the individual traits. 
A selective way of birds that reacts to their neighbors is one of the main factors characterizing the collective behaviors. 
Individual birds can react differently depending on their local environment during the collective decision-making process, and these variable reactions can be a source of complex spatiotemporal flocking dynamics. 
Here, we extend the deterministic Cucker-Smale model by including the individual's reaction to neighbors' acceleration where the reaction time depends on the local state of polarity. 
Simulation results show that the adaptive reaction of individuals induces the collective response of the flock. Birds are not frozen in a complete synchronization but remain sensitive to perturbations coming from environments.
We confirm that the adaptivity of the reaction also generates natural fluctuations of orientation and speed, both of which are indeed scale-free as experimentally reported. 
This work may provide essential insight in designing resilient systems of many active agents working in complex, unpredictable environments. 

\end{abstract}

\maketitle

%\tableofcontents

\section{Introduction}

%% [1]
%1
Various behavioral patterns of natural flocks in open space arise in response to internal or external perturbations \cite{couzin2007collective,cavagna2014bird}. 
%2
Birds constantly adjust their behaviors according to their neighbors' states, synchronizing their dynamic states without losing coherence. 
%3 = 1+2 정리
The collective behaviors of birds can be characterized by social interactions between them \cite{ballerini2008empirical,carere2009aerial,attanasi2015emergence,couzin2018synchronization,wang2019collective,ren2018stable}. 
%% 모델
%4 모델 관심
Many individual-based models have been proposed to analyze key aspects of collective motions, for example, bounded group formation, velocity alignment and speed control \cite{acebr2005kuramoto,hemelrijk2011some,gerlee2017impact,mora2016local}. 
%5 비체크 모델
The classical models like Vicsek (discrete-time) and Cucker-Smale (continuous-time) models have focused on the emergence of velocity synchronization, assuming that the speed is constant. They have used simple rules of the heading alignment: birds change their headings to be aligned with the average of those of their neighbors, and their behavioral changes occur spontaneously without delay \cite{vicsek1995novel,cucker2007on,cucker2007emergent}. 
%6 비체크 모델 요동
Especially, the stochastic noise in Vicsek's model represents behavioral errors (fluctuations) of birds in following the rules and leads to a dynamic phase transition.  
%7 모델 패턴
The common dynamic patterns in ordered phases are mostly simple geometries like straight lines, compact circles, or ellipses, depending on the strength of a pairwise potential for the group formation \cite{orsogna2006self}.
Interestingly, in addition to noise, a fixed delay in social interactions can cause behavioral changes \cite{erban2016cucker}. If a fixed delay in velocity is short, there is no qualitative difference occurring in the collective behavior \cite{liu2014flocking}; If a fixed delay in velocity becomes long, it can cause bistability and induce kinetic phase transitions, for example, from milling to clumped rotating states \cite{hindes2020unstable,forgoston2008delay}.

%% (요동) [2]
%1 정렬상태 문제
%
Flocking solutions obtained from classical models without noise and delays maintain a completely ordered state which is stable with respect to perturbations: perturbed individuals instantly return to the synchronized state, and the flock remains non-responsive to their environments in time \cite{okubo2001diffusion}.
In some cases, such complete synchronization may not be desirable for a natural flock, since perturbations can contain crucial information for survival, e.g., predators' attacks or imminent changes of directions for food, which may instantaneously disappear in a non-responsive state. 
%4깰 방법이 없다.
In these classical models, without introducing any additional forces breaking such a synchronized formation, it would be impossible for a flock by itself to shift from a ``lifeless" steady state to a state responsive to its environment.  
% %% 요동 
%6 실제 새 요동의 특징 7-8-9
In complex systems, it is often natural fluctuations that continuously drive a system to a ``desirable" state \cite{chandler1987introduction}. 
It has been reported that the behavioral fluctuations of actual birds are distinct from stochastic noise. 
% %7 코릴레이션 중요
% A flocking behaviour of real birds is well characterized by the correlation in their dynamics.
%8 무척도
The correlations of the fluctuations in the orientation and the speed, respectively, are proportional to the group size. This implies that correlations are scale-free and therefore maximize the speed of the information transfer across the flock \cite{cavagna2010scale,attanasi2014information}. 
%9 임계계
A natural flock behaves as a critical system in that birds keep ready to maximally respond to environmental perturbations \cite{mora2011biological,hemelrijk2015scale}.
%criticality \cite{chuang2016swarming,topaz2012locust} hemelrijk2015scale 
% 모델계
%
However, the classical flocking models have a limited capability to describe these properties: the self-propulsion speed is set constant; the density of a flock is spatio-temporally uniform; lastly, the noise only serves as an analogy for the temperature and is therefore random and uncorrelated in time.
%10 행동 패턴 무정형
% It is also notable that the density of a natural flock can be spatially non-uniform, creating complex and amorphous dynamic patterns. 

%[3] 문제 설정 
%1
An interesting question would be what is the underlying inter-individual coordination mechanism for a living flock not to be stuck in synchronization but to keep stay in a sensitive state while maintaining its polarity \cite{couzin2007collective,cavagna2010scale,lukeman2010inferring,gautrais2012deciphering}.
%2
There have been few flocking models that can reproduce the scale-free correlations of fluctuations in both the velocity and the speed as observed in experiments \cite{topaz2012locust,chuang2016swarming}. Recently, a statistical model called the maximum entropy model has been developed by Bialek {\it et al.} to describe such scale-free correlations  \cite{bialek2012statistical,bialek2014social}. They measure the local correlations and the variances of the speed from the field data and use them directly as inputs of their model. Although their model successfully predicts the flocking dynamics of a real flock, for a deeper understanding of generic flocking behaviors, we still need to investigate a rule-based mechanism that individual birds may follow.
%3

%% [4]
%1 가속도 반응
Birds should be able to detect others' velocity changes and accelerate/decelerate accordingly in order to avoid collisions or keep the group's cohesion tight \cite{cavagna2010scale,gerlee2017impact,karamouzas2014universal}. 
%2 실험 가속도 반응
Experimental data shows that during the sharp turning of a flock, the acceleration profile of each bird with time is similar except a time shift. This implies that each bird responds to others' acceleration with a delay. Indeed, when the speed of the group is not uniformly constant, a finite reaction time is required for each bird to accelerate/decelerate in response to neighbors' speed changes \cite{nagy2010hierarchical}. 
Compared to the delay in the velocity response \cite{liu2014flocking,erban2016cucker,hindes2020unstable}, as we will see through this work, the delay in the acceleration response can yield a substantial difference in flocking solutions. 
%3 모델
%Importance of the acceleration response has been recognized and implemented in an individual based model to investigate complex collective motions
%4 현 상황
However, partly due to the typical assumption of a constant speed that is widely used, there have been a relatively small number of studies recognizing the importance of acceleration synchronization in individual-based models. 
%5 관련 연구 1
Among few studies, Szabo {\it et al.} proposed a generalized Vicsek model coupled with acceleration with a short reaction delay which serves as a separate perception mechanism \cite{szabo2009transitions}.
%6 연구 결과
They found that flocks undergo a novel order-disorder phase transition depending on a value of the strategy parameter that determines the relative contribution of the acceleration synchronization and the velocity synchronization. 
%13 
The flocking dynamics of their model become noisy and disordered as the strategy parameter gradually increases with a stochastic noise magnitude fixed.
%10 그 문제 원인
Although the effect of the reaction delay in acceleration was not explicitly discussed, one can guess that a short reaction delay ($\Delta t$) implemented in their model is responsible for the novel phase transition. It is because we notice that for a continuous time model, the acceleration coupling without any delay simply induces instability in flocking solutions.

%밀도 고정 [5]
%1 정보전파속도와 오더
Recently, it has been found that the collective response is related to an ordering state of a flock. 
A previous experimental study by Attanasi {\it et al.} showed that during a circular turning with a constant speed, localized perturbations propagate without damping from bird to bird across the whole group and the speed depends on the global polarization \cite{attanasi2014information, mora2016local}. 
%the square of the speed for information transfer 
They also developed a related model including behavioral inertia (in terms of a phase angle) and conservation laws under the assumption of the uniform polarity \cite{cavagna2015flocking,attanasi2014information}.
%3밀도 분포가 있을 때
It is reasonable, for more complex flocking dynamics, to consider spatially-varying local orders at an individual level. This can naturally induce non-uniform reactions among the flock.  
%4반응 시간 
In other words, the local speed of information transfer (i.e., the local reaction time of an individual) may differ from bird to bird depending on a local state. 

%% [6] 
%1 모델
Here, we consider a generalized Cucker-Smale model to study the effect of an individual's reaction to acceleration with a delay on collective behaviors \cite{cucker2007on,cucker2007emergent,agueh2011analysis}. 
%2 강조
Non-uniform reaction times of birds to acceleration are considered as an as-yet-unexplored property of collective behaviors  \cite{chat,insp,kama2016,kama2018,jin}. 
%3 모델 세부
We also include social interactions commonly found in classical models: velocity synchronization, speed control by drag, and group formation in open space through a pairwise potential \cite{cucker2007emergent,niwa1994self}. 
%4 적응적 반응 고려
Based on the fact that the speed of the information transfer depends on the group polarity, we assume that an individual's delay to its neighbor's acceleration depends on the local polarity. 
% %5 이전 모델 비교
% Unlike Szabo's model where the reaction delay is constant in space and time independent of individuals, reaction delays are adaptive in that they are proportional to local polarities and thus non-uniform in space and time. 
%
%6 행동 결과 
Simulation results show that birds form a flock with a collective response and remain sensitive to perturbations of others without losing its polarity. This unique behavior is caused by the interplay between the velocity alignment and the adaptive acceleration synchronization. 
%7 파라미터 
We also introduce a parameter, $\kappa$, called a sensitivity constant that controls the level of responsiveness of birds and study its effect on the collective dynamics. 
%
%10 요동 결과 
Even if we do not include any stochastic noise, it is shown that there are natural fluctuations in the solution, intrinsically generated with the adaptive acceleration delay. 
%12
Most importantly, we found that these fluctuations are random but not independent: they are correlated and scale-free in both velocity and speed, which is the signature of natural flocks with great adaptability and resilience.

\section{Model}
%%[7] 모델 도입. 
We develop a deterministic flocking model coupled with the acceleration response of birds with a delay. 
Our model is based on the Cucker-Smale type alignment rule of each bird that includes smoothly decaying adjacent function with distance: Each bird reorients its heading according to the weighted average orientation of neighbor birds within the radius $r_0$ \cite{cucker2007on,cucker2007emergent,agueh2011analysis}.
We use a Euclidean metric distance between birds $i$ and $j$ as $d_{ij}=|x_i-x_j|$. While there are other types of distance based on the visibility or the network topology, the metric distance turned out to be a useful tool to investigate animals' collective behavior \cite{ballerini2008interaction,haskovec2013flocking,bastien2020model,brown2021information}. 
In our model, birds also synchronize their accelerations with neighbors' average accelerations as they do their headings with neighbors' averaged directions. 
We incorporate a time delay with the acceleration response, considering that reaction delays in velocity and acceleration may differ with separate time scales.
We also include two additional social forces that are commonly adopted for the study of the flocking: (1) a drag-like force, leading to the natural reference speed $v_0$ in the absence of other forces and (2) a pairwise potential, which has the long-ranged attraction to hold a group in an open space and the short-ranged repulsion to adjust its separation distance. 
We consider a flock of $N$ birds in a two-dimensional open space. A state of bird $i$ for $i=1,\cdots,N$ at time $t$ is described by the position $\textbf{x}_i(t)$ and velocity $\textbf{v}_i(t)$. 
%%[8] 수식1 첫번째 항
We describe the flocking dynamics using the deterministic Cucker-Smale model: 
\begin{eqnarray}\label{maineq1}
\frac{d\textbf{x}_i}{dt} &=& \textbf{v}_i \\ 
\frac{d\textbf{v}_i}{dt} &=& \sum_{j=1}^{N}J_{ij}(\textbf{v}_j-\textbf{v}_i)
+\sum_{j=1}^{N}I_{ij}\tilde{\textbf{a}}_{j}\nonumber + \psi(\textbf{v}_i) - \nabla_i \phi(\textbf{x}_i) 
\end{eqnarray} 
where the gradient $\nabla_i=\partial / \partial \textbf{x}_i$. The first two terms on the right-hand side define the self-propulsion mechanisms of bird $i$ concerning the neighbors' velocity and change of velocity, i.e., acceleration. 
\textcolor{black}{To consider the variable influence of neighbors depending on their metric distances, we calculate weighted averages using the interaction matrices $J_{ij}$ and $I_{ij}$} that measure the influence of bird $j$ on $i$. Since the influence monotonically decays with the distance $d_{ij}$, we take $J_{ij} = K_v g(d_{ij})$ and $I_{ij}=K_a g(\tilde{d}_{ij})$, where $g(y)=\frac{1}{(1+y^2)^2}$, and $K_v$ and $K_a$ is the interaction strength for the velocity and the acceleration, respectively \cite{cucker2007on,cucker2007emergent}. 
Here the acceleration of $j$, $\tilde{\textbf{a}}_j$, and the distance between bird $i$ and $j$, $\tilde{d}_{ij}$, are computed with a delay at $\tilde{t}_i=t-\tau_i$. 

\textcolor{black}{At the moment when a bird changes its heading according to the velocity rule, it probably does not recognize the neighbors' acceleration/deceleration, since mimicking others' acceleration requires information regarding the difference between the velocities of neighbors in time. Here we assume the separate time scale between velocity (often called a soft mode) and acceleration (a stiff mode) responses.} Unlike instantaneous alignment of each bird mimicking the neighbor's heading, it takes a finite time of $\tau_i$ to react to the neighbor's momentum change.
We point out that, although the velocity $\textbf{v}_i$ appears in a relative form ($\textbf{v}_j-\textbf{v}_i$) in the equation and the acceleration $\textbf{a}_i$ does not, \textcolor{black}{the both relative/absolute forms play the same role to attain the average velocity and the average acceleration of its neighbors', respectively, at the steady state (i.e., $d\textbf{v}_i/dt=0$). The different forms are simply due to the algebraic structure of the equation where we have the acceleration on the left-hand side of the equation of motion.}

%%[9] 수식 설명
%두번째 항
Motivated by the fact that the speed of information transfer across the entire group depends on the group order \cite{attanasi2014information}, we assume that the reaction time $\tau_i$ of bird $i$ depends on the local order of $R_i$ :
\begin{equation}\label{tau1}
\tau_i = \kappa \left( 1- R_i \right)
\end{equation} 
where $\tau_{i}$ is the time taken for bird $i$ to change its behavior in response to the neighbors' changes. 
Here, $\kappa$ is a coefficient related to the reaction sensitivity. If $\kappa$ is large, the reaction of bird $i$ is slow, so we can say the bird is insensitive to its neighbors' changes. On the other hand, if $\kappa$ is small, the reaction is fast and thus we can say the bird is sensitive to perturbations. In our simulations, we take the intermediate value of $\kappa$ between the two extremes: if a reaction delay is too long ($\kappa$ is too large), it would be non-physical, while if a reaction delay is too short ($\kappa$ is too small), the dynamical system becomes unstable, merely giving disordered states.  
$R_i$ is the local order of the neighbors of bird $i$:
\begin{equation}
R_i=\frac{1}{1+\sigma_i}
\end{equation}
where $ \sigma_i = c_0 \times \textrm{the variance of }  \{\textbf{v}_j | \textrm{ for all }j \textrm{ where } d_{ij}\leq r_0$\} and $c_0$ is a constant. 
Note that $R_i$ becomes 1 ($\sigma_i=0$) in a perfect alignment and approaches zero as the local variance increases ($\sigma_i\rightarrow \infty$). 
The group order is computed as the average of the local orders \cite{acebr2005kuramoto,okeeffe2017oscillators,levis2019activity}
\begin{equation}
R=\frac{1}{N}\sum_i^N R_i
\end{equation}
When a flock is in a perfect local alignment ($R_i=1$), its reaction becomes instantaneous ($\tau_i = 0$). 
%
% This increases errors (in the form of noise) in birds' obeying behavioral rules. 
%
When the group's local alignment is completely random ($R_i=0$), the reaction time is equal to be $\tau_i=\kappa$. As the magnitude of $\tau_i$ increases in between, a bird's reaction becomes slow (and thus we can say they are less sensitive to others' momentum changes). 
    We will show that such state-dependent reactions of individuals give remarkable flexibility and resilience to the dynamics of the flock. 
    Our model is different from an extended Vicsek's model with the acceleration coupling in that the bird's reaction is not instantaneous but is adaptively delayed \cite{szabo2009transitions}.

%%[10] 수식 설명
%수식 3번째 항 
The third term in the equation is the speed control: The drag force is exerted on a bird if its speed becomes away from the natural reference group speed $v_0$:  
\begin{equation}
   \psi(\textbf{v}_i)=\alpha(v_0^2-|\textbf{v}_i|^2)\textbf{v}_i 
\end{equation}
Note that this allows the variable speed of a bird, whereas in the classical Vicsek's model the speed is constant $v_0$ in time for all $i$.
% 수식 4번째 항
To form a bounded group and also to avoid the crash of birds, a pair potential $\phi$ is introduced. It combines the Morse potential for exponentially decaying short-range repulsion and the wall potential for gently increasing long-range attraction with a degree three  \cite{gerlee2017impact,orsogna2006self}. \textcolor{black}{The cubic function is chosen since it gives a proper characteristic scale for the flock size in the simulations. A function with a higher order such as 4 or 5 provides the stronger force of the group formation than the cubic function, possibly limiting the variable flocking dynamic in free space.}
\begin{equation}\label{phi}
\phi(\textbf{x}_{i})=\sum_{j=1}^N \left(C_r {e}^{-|\textbf{x}_j-\textbf{x}_i| / l_r}\right) + C_a |\textbf{x}_j-\textbf{x}_i|^3. 
\end{equation}
Here $l_r$ is an effective distance of the repulsion, $C_r$ is the strength of the short-range repulsion, and $C_a$ is an effective distance of the attraction. In the case $C_a=0$, the potential is governed by repulsive behavior and birds tend to disperse into the entire volume, corresponding to the H-stable regime V in ref \cite{orsogna2006self}. As $C_a$ increases, the group of birds tends to be organized into a structure with a well-defined inter-individual distance as in the case of H-stable regime VI in ref \cite{orsogna2006self}. \textcolor{black}{Note that the normalized equations of Eq~(\ref{maineq1}) by the characteristic scales have the same forms as the original equations. Thus we use dimensionless variables and solve the normalized equations in the following unless otherwise stated.}

%%[11] 파라미터 U and G
The group dynamics are measured by using the two other macroscopic parameters $U$ and $G$, the group speed and the group size, respectively. A group speed $U$ is the average speed of the flock.
\begin{equation}
U=\frac{1}{N} \left |\sum^N_{i}\textbf{v}_i \right |
\end{equation} 
A group size $G$ is the average distance between an arbitrary bird and the center of mass of a flock.
\begin{equation}\label{G}
G=\frac{1}{N}\sum_{i=1}^N |\textbf{x}_j-\textbf{x}_0|,
\end{equation}
where $$\textbf{x}_0=\frac{1}{N}\sum_i^N {\textbf{x}_i}$$ is the center of mass of a flock.
    Note that, instead of directly measuring the occupied area by birds in space, the group size $G$ defined by the distances between birds provides a generalized way to measure a dynamical size of a flock which could be amorphous or sometimes abruptly splitting. 

\begin{figure}[!h]
            \includegraphics[width=0.95\textwidth]{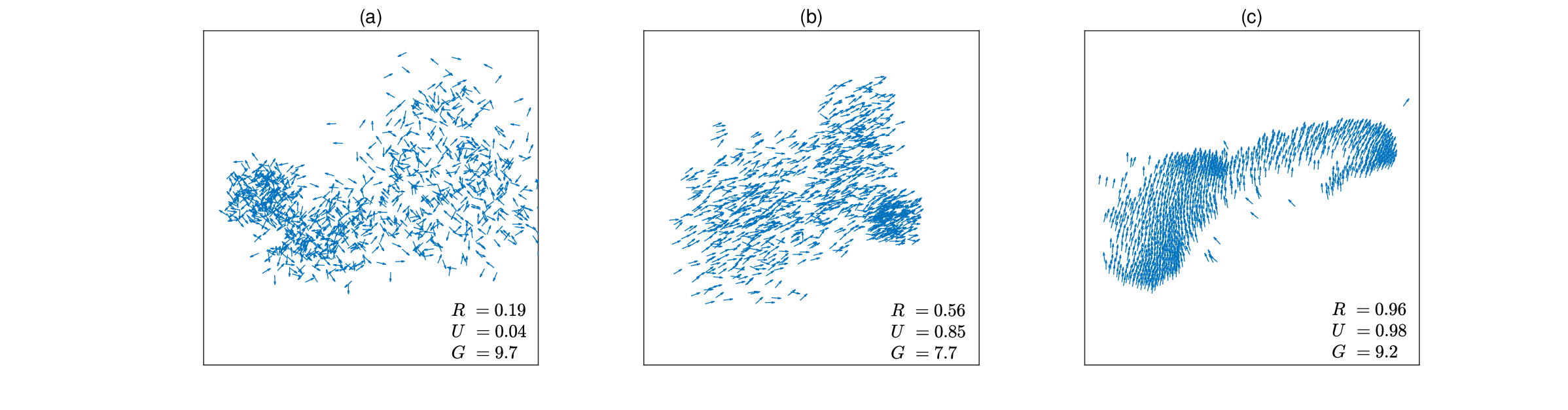}
            \caption{{\bf Instantaneous velocity fields when $N=1000$ }}
            \label{fig1}
            \end{figure}

%%[12]
%Fig 1 속도 필드 
As an example for a flock formation, we illustrate instantaneous velocity fields and the corresponding macroscopic parameters in Fig. \ref{fig1}. The number of birds is set as $N=1000$ in each case, and the velocity vectors are scaled to one for clarity. 
In the configuration of (a), the group order is $R=0.19$ and the directions of the velocities are randomly distributed. The corresponding group speed becomes close to zero as $U=0.04$. In the configuration of (b), $R=0.56$ and birds form a more coherent group. In (c), the order parameter is close to 1, i.e., $R=0.96$ and it clearly shows that birds are in a synchronized state moving in the same direction with the group speed of $U=0.98$, which is much faster than the case in (a). The corresponding value of the group size $G$ is also indicated in each panel in Fig. \ref{fig1}.

% Results and Discussion can be combined.
\section*{Results and Discussion}
%
%% Setting 
%%[13]
We investigate the collective dynamics of birds when $N=1000$. The number of birds is chosen to display large scale patterns but is limited by the computation cost. The collective behaviors of the simulations with the larger $N$ are qualitatively similar. Our main focus is on the effects of acceleration interaction in terms of $K_a$ and $\kappa$ while fixing other constants. \textcolor{black}{All parameters are presented in dimensionless numbers normalized by the length scale of $l_{scale}=1$ m/s and the time scale of $t_{scale}=0.1$ sec. This leads to the velocity scale of $v_{scale}=10$ m/s}: $r_0=0.6$, $K_v=0.1$, $v_0=1$, $c_0=10$, $\alpha=0.2$, $C_r=1.5$, $l_r=0.05$, and $C_a = 1.5\times 10^{-7}$. \textcolor{black}{Note that all results presented in the following are dimensionless ones except the time $t$ in plots, which is in seconds (i.e., 1 sec = 10$t_{scale}$).} 
We set the coefficients of $C_r$, $l_r$ and $C_a$ in the simulations so that the corresponding group formation falls on the H-stable regime VI in ref. \cite{orsogna2006self}. Using the typical constants, the natural length scale becomes $\tilde{r}_p\approx 1$ which corresponds to the minimum of the pair-potential, satisfying $C_r/(3C_al_r) = \tilde{r}_p^2 e^{{r}_p/l_r}$.

\begin{figure}[!h]
            \centering 
            \includegraphics[width=0.95\textwidth]{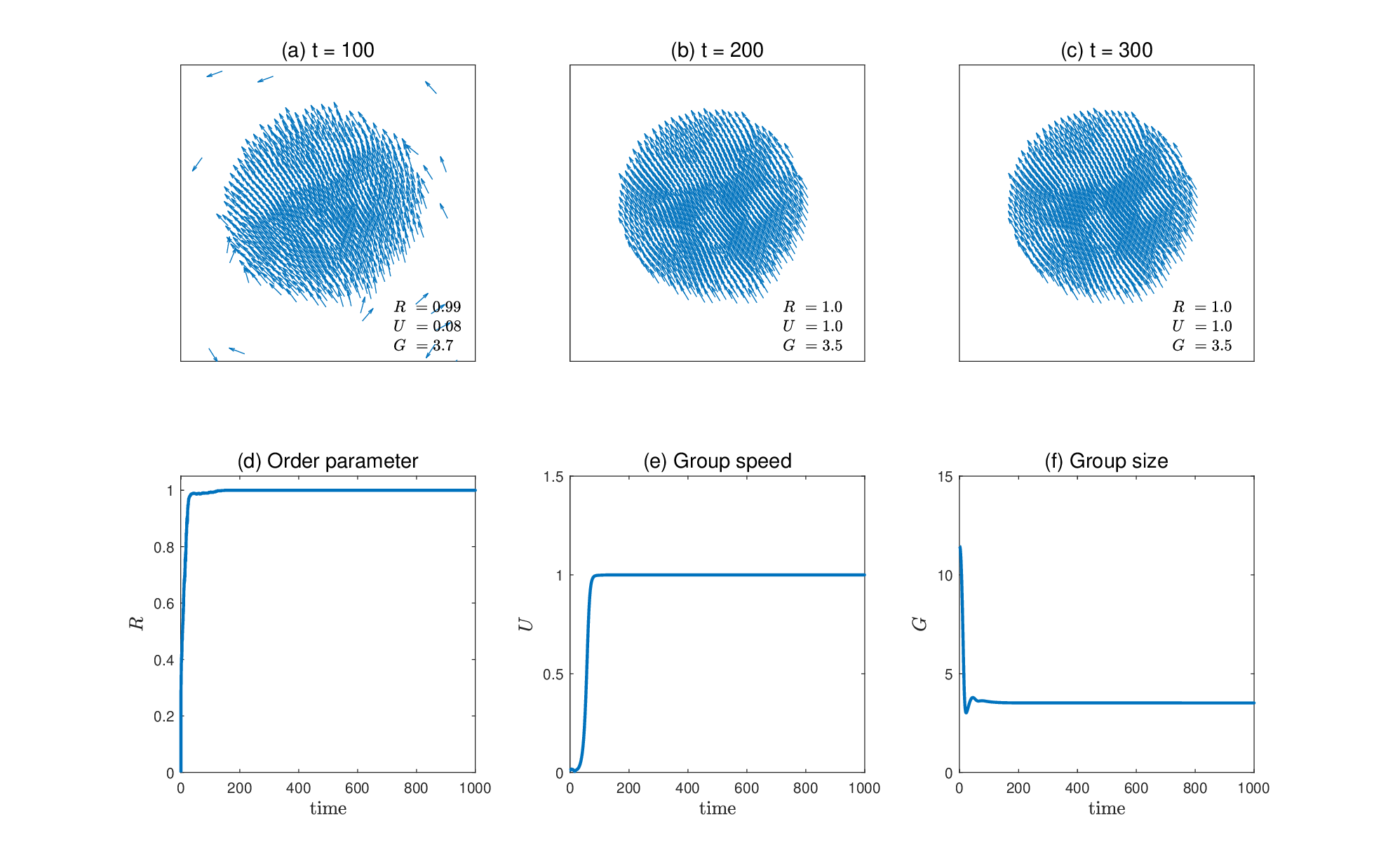}
            \caption{{\bf Time evolution of a flock without the coupling of acceleration.}
            We use Eq (\ref{maineq1}) when $K_a=0$. The number of birds is $N=1000$ and the vectors are scaled to one for clarity. The first row is the instantaneous velocity fields of a flock at (a) $t=100$ sec, (b) $t=200$ sec, and (c) $t=300$ sec. The second row is the dynamic state variables of (d) $R$, (e) $U$, and (f) $G$ as a function of time.}
            \label{fig2}
            \end{figure}

%% Figure 2 
%%[14]
% 
We first solve normalized equation of Eq~(\ref{maineq1}) without the acceleration coupling at $K_a=0$. We assume that the initial positions and velocities of birds are randomly distributed. The final simulation time is $t_f=1000$ sec, which we ensure is sufficiently larger than the transient time scale. 
%
% 그래프 top
Figures ~\ref{fig2}(a)--(c) show the snapshots of the velocity fields at different times, (a) $t=100$ sec, (b) $t=200$ sec, and (c) $t=300$ sec. 
    % 결과 top
    We observe that the system rapidly converges to a synchronized state. The birds form a circular shape and perform a rectilinear movement while their headings are perfectly aligned in time.
    %그래프 down
The flocking dynamics are also quantitatively shown in terms of macroscopic variables in Figs \ref{fig2}(d)--(f). % 결과 down
    The flock rapidly converges to the synchronized values of $R=1$ and $U=1$, and the size of the group remains constant. 
        Since our model is deterministic, in the case without the acceleration term, the solution corresponds to the perfectly ordered state of the classical Vicsek model. 
        Once the group achieves the synchronized state which is stable, the flock is non-responsive to any perturbations. 
    %% 속도 딜레이 혹은 가속도 상수 딜레이 디스커션 넣을 것 
%
       \textcolor{black}{One the other hand, if there is only an acceleration rule with $K_v=0$ and $K_a\neq0$, the dynamic state of a flock in directional responses is overall random and noisy at a given $\kappa$, since acceleration rule is not directly related to the directional alignment (See the time evolution of the order $R$ when $K_v=0$ in the Supplementary Information). The time-averaged group order tends to decrease (more disordered and noisier) as $K_a$ increases at a given $\kappa$ due to the locality of the acceleration responses (see Fig. \ref{fig5}(a)). Note that the order also decreases as $\kappa$ decreases (i.e., response time decreases) at a given $K_a$ as discussed in Fig. \ref{fig7}(a).} 

\begin{figure}[!h]
            \centering
            \includegraphics[width=0.95\textwidth]{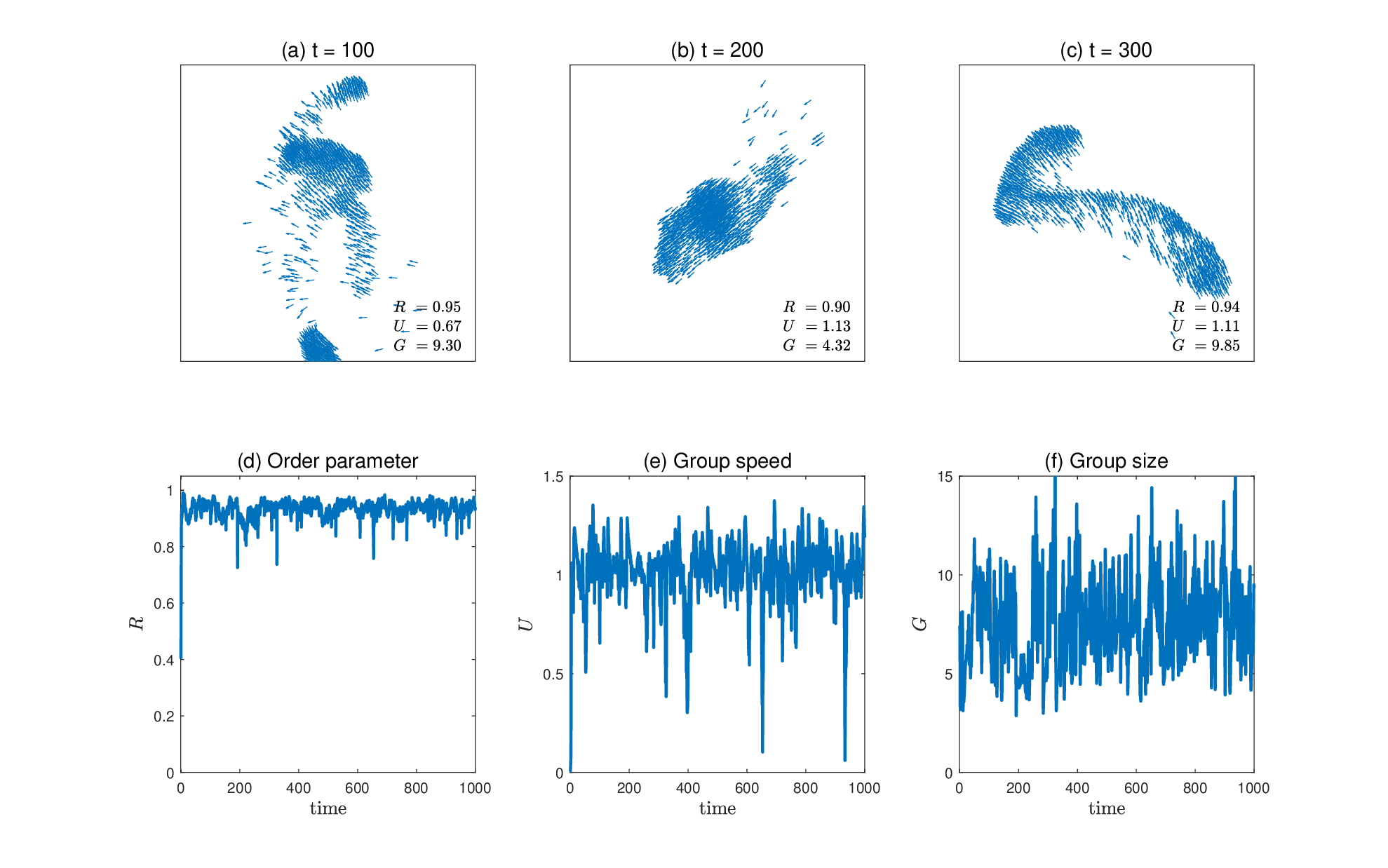}
            \caption{{\bf Time evolution of a flock with the coupling of acceleration.} We use Eq (\ref{maineq1}) when $K_a=0.12$ and $\kappa=800$ (equivalently 80 s), and other parameters are the same as in Fig. \ref{fig2}. Velocity fields of a flock at (a) $t=100$ sec, (b) $t=200$ sec, and (c) $t=300$ sec. Dynamic state variables of (d) $R$, (e) $U$, and (f) $G$ as a function of time.}
            \label{fig3}
            \end{figure}

%% Figure 3
%% 그림 top
Before we discuss the roles of the parameters $K_a$ and $\kappa$, respectively, we show representative behaviors of a flock when $K_a=0.12$ and $\kappa=800$ \textcolor{black}{(equivalently 80 sec)}. We assume that the positions and velocities of birds are initially randomly-distributed and this initial configuration is used as a template configuration for those at the negative time $t<0$ for the delay term. \textcolor{black}{To deal with the non-uniform delay, we simply rounded the delay time to the nearest mesh point to avoid excessive computation involved in the long-term integration of the large system ($N \geq 1000$, $t \geq 10000$). The equation has been integrated by the backward finite difference method with the order three accuracy. As long as the acceleration has bounded variation, which we can reasonably assume for bird flocks, such numerical integration has the error of the magnitude $O(\Delta t)$ where $\Delta t$ is the mesh size. We numerically confirmed that the solution converges well as $\Delta t \rightarrow 0$.}
$K_a$ determines the relative contribution of the acceleration synchronization compared to the strength of alignment $K_v$, while $\kappa$ determines its sensitivity in responding others' accelerations through reaction delays. 
Figures ~\ref{fig3}(a)--(c) are the snapshots of the velocity fields at different times, (a) $t=100$ sec, (b) $t=200$ sec, and (c) $t=300$ sec. 
%%
    %%[15] 그림 3 개요. 
    With the interplay between the acceleration synchronization and the velocity alignment, the flock shows markedly coherent dynamics with variability in shapes and densities. 
    At a given time, one part is in a lower density and another part is less ordered, displaying the locality that could be commonly observed in a natural flock. (See Supplementary Videos for the collective dynamics with the coupled acceleration.) 
    %% 결과 top

%% 그림 bottom
The quantitative feature of the group order is shown in Fig. \ref{fig3}(d). 
    Due to the coupling of the ordering and the reaction time (sensitivity), the behavior of the solution is dramatically different from the previous case with $K_a=0$ which has a stable fixed-point solution at the perfectly ordered state. 
    The complete synchronization is now an unstable fixed-point solution, since the bird's reaction becomes instantaneous ($\tau_i=$0) at $R=1$. \textcolor{black}{Since the propagation speed is inversely proportional to $\tau$, the shorter $\tau$ induces the faster propagation of the small perturbations present in the system (see the stability analysis in Appendix A). Due to the coupling between the response time and the local order, the larger magnitude of fluctuations can be induced at higher $R$. 
    If the system is ideally located at $R=1$ from the beginning, it stays at that point for all $t$ since our model is deterministic and there is no stochastic noise. 
    However, if a flock is not initially in a perfect alignment}, the value of $R$ can be maintained near one but never completely converges to one like the previous case with $K_a=0$. 
    % 결과 bottom
     %
    %
        It can be partly understood from \textcolor{black}{the fact that the order (proportional to $K_v$) and the sensitivity (inversely proportional to the reaction delay $\tau$) are coupled as a negative feedback loop. On one hand, the highly ordered states induce the more sensitive responses. On the other hand, the highly sensitive responses prevent a flock from being highly ordered. This mechanism is possible through the delay time that depends on the local order. This negative loop is the characteristic feature of the rich dynamics based on the interplay between $K_v$ and $K_a$ terms.} 

            Whereas a spontaneous reaction time at an ordered state can maximize the speed of information propagation across the entire group, the flock becomes inevitably unstable: a bird is subject to magnify neighbors' erroneous behaviours of fluctuations, leading to a break of directional synchronization. 
            %% 3 d discussion 2 
            Once the reaction time gets longer at a lowered polarity, the state tends to be ordered back from the first rule of their headings. 
            %% 3 d discussion 3
            This flocking behavior indicates that the trade-off, gaining the group sensitivity at the expense of the group order, may be crucial for a real flock: a flock can achieve both polarity and cohesion (sensitive to others' behavioral change) at an optimized level as a critical system. 
            %%
            %
%% 그림 bottom e and f 
In Figs~\ref{fig3}(e) and (f), the influence of the acceleration coupling on the group speed $U$ and the group size $G$ are also shown, respectively. 
    Although our model is deterministic without a stochastic noise, natural fluctuations in $U$ and $G$ are presented. \textcolor{black}{A large magnitude fluctuations caused by a short delay near $R=1$} destroys the group polarity, leading to morphological changes or temporal ruptures. 
    The statistical analyses of the fluctuations in orientation and speed are performed using correlations functions discussed in Fig~\ref{fig8}.
            To further obtain insight into these behaviors, we provide the simplified linear instability analysis in Appendix A. 
               Note that in our model the delay is only accounted for in the acceleration due to the assumed separate time scales of reactions. 
        Our preliminary simulation results show that even if an adaptive delay is included in velocity, it does not qualitatively influence collective behaviors at the given parameters.
        As we previously discussed, if we have a fixed delay in acceleration response, its collective behaviors are qualitatively similar to those from the Szabo's model \cite{szabo2009transitions}. \textcolor{black}{See the result comparison of these cases provided in the appendix B.}

    %

    % %% 결과 bottom e and f velocity and location 
    % The essential factor for the unique behaviors is the competition between the velocity synchronization for the ordering and the acceleration synchronization for the susceptibility. 
    %as a critical system
    %%comparison
 
        %comparison results

\begin{figure}[!h]
            \centering
            \includegraphics[width=0.95\textwidth]{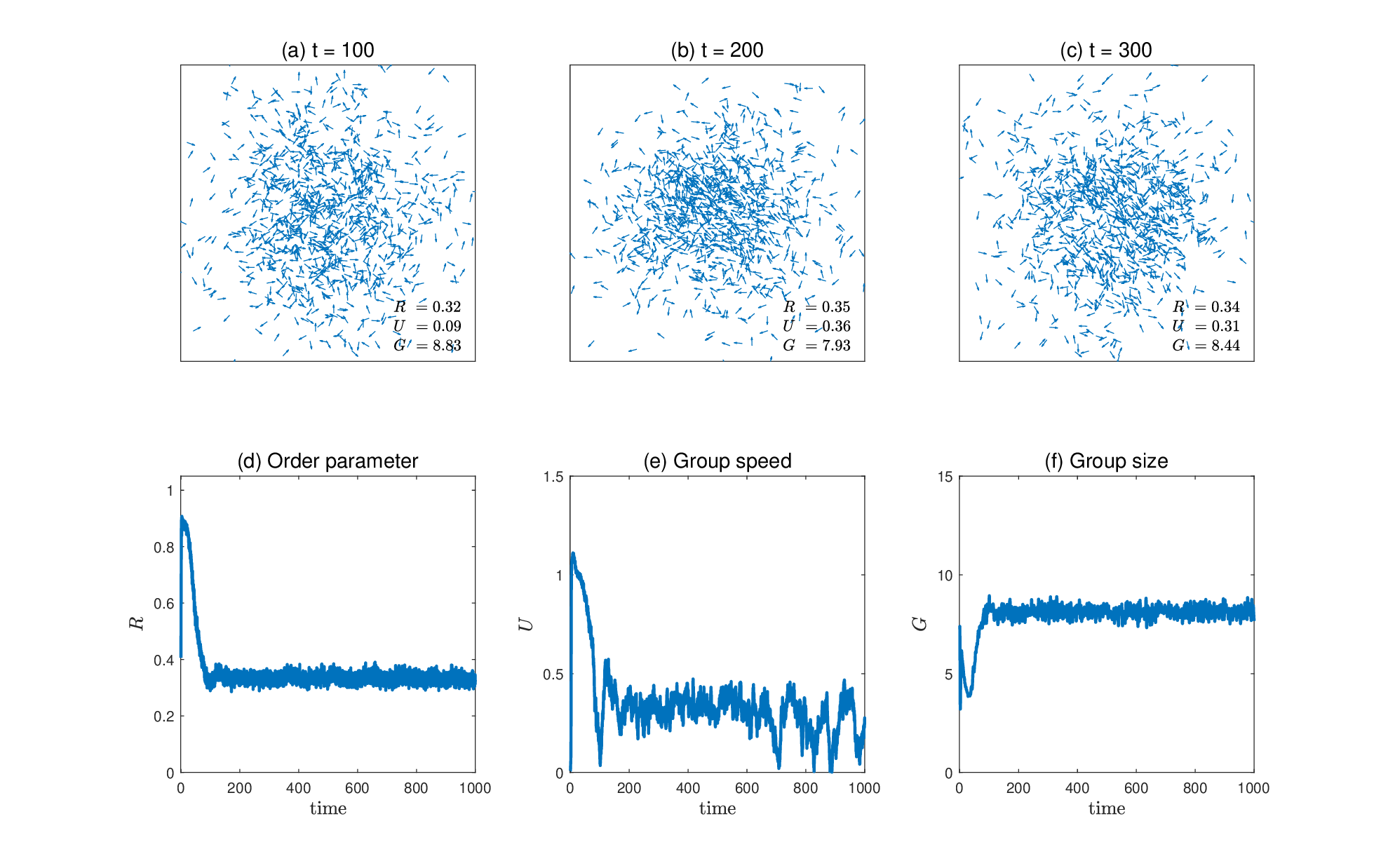}
            \caption{{\bf Time evolution of a flock with the coupling of acceleration.} We use Eq. (\ref{maineq1}) when $K_a=0.24$ and other parameters are the same as in Fig. \ref{fig3}. Velocity fields of a flock at (a) $t=100$ sec, (b) $t=200$ sec, and (c) $t=300$ sec. Dynamic state variables of (d) $R$, (e) $U$, and (f) $G$ as a function of time.}
            \label{fig4}
            \end{figure}

\begin{figure}[!h]
            \centering
            \includegraphics[width=0.95\textwidth]{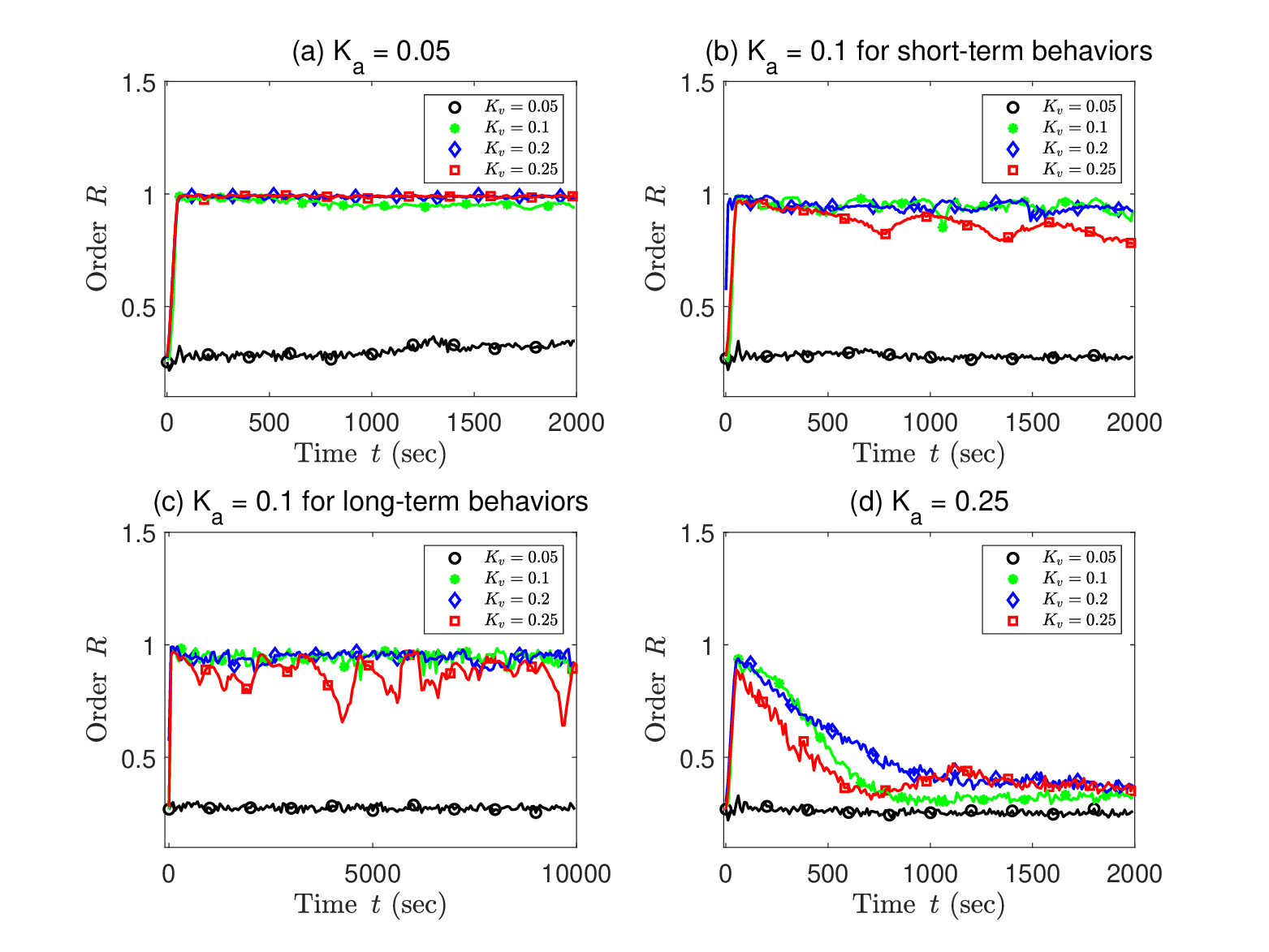}
            \caption{{\bf The effect of $K_v$ on the time-evolution of $R$ for various values of $K_a$ when $\kappa=800$}}
            \label{fig5}
            \end{figure}

We now consider the effect of the doubled strength of the acceleration synchronization with the value of sensitivity constant fixed. 
Figures~\ref{fig4}(a)--(c) show snapshots at $t=$\textcolor{black}{100, 200}, and 300 seconds from a simulation with the same parameters as in Fig.~\ref{fig3} except the value of $K_a=0.24$. 
    We observe more random and less coherent aggregation, where the effect of the heading alignment is less pronounced than that of the acceleration synchronization. 
    After a transient time passes near at $t=100$ sec, the group order falls down to below $R=0.4$ and this value is maintained with time (Fig.~\ref{fig4}(d)).
        % This can be interpreted as birds are less responsive to other's behavioral changes with the elongated reaction delays. 
    Also, $U$ decreases and $G$ increases according to the lowered group order. 
    %
    % Note that the magnitude of fluctuations is similar to the previous case in Fig.~\ref{fig3} since the sensitivity constant $\kappa$ is the same. (See the further discussion in Fig~\ref{fig8}.)
    %

% Figure 4

\textcolor{black}{As $K_v$ increases for various values of $K_a$, the time-averaged order increases toward one (more ordered), as similarly reported in \cite{szabo2009transitions}. However, there is another aspect to the flocking dynamics that we like to highlight.  
In Fig. \ref{fig5} we show that there is the negative feedback loop between the group order and the sensibility through the reaction time, leading to short-term regular or long-term irregular oscillating motions depending on the strength of $K_a$. 
For a small value of $K_a=0.05$ (Fig. \ref{fig5}(a)), the flock is in a velocity-dominant regime and the order of the flock increases with $K_v$.  
For a large value of $K_a=0.25$ (Fig. \ref{fig5}(d)), the flock is in the acceleration-dominant regime and the order is less than 0.5 even with a large value of $K_v$.
For the intermediate value of $K_a=0.1$ (Fig. \ref{fig5}(b) or (c)), the effects of $K_v$ and $K_a$ are competing, making interesting collective behaviors. At the large value of $K_v=0.2$ (blue diamonds) or 0.25 (red squares), the fluctuations appear, which become apparent in the long-terms behaviors in Fig. \ref{fig5}(c). They are more irregular and long-lasting compared to the case of $K_a=0.05$ in Fig. \ref{fig5}(a) or $K_a=0.25$ in Fig. \ref{fig5}(d).}

%%New Kv study quantatative study
% Figure new

\begin{figure}[!h]
            \centering
            \includegraphics[width=0.99\textwidth]{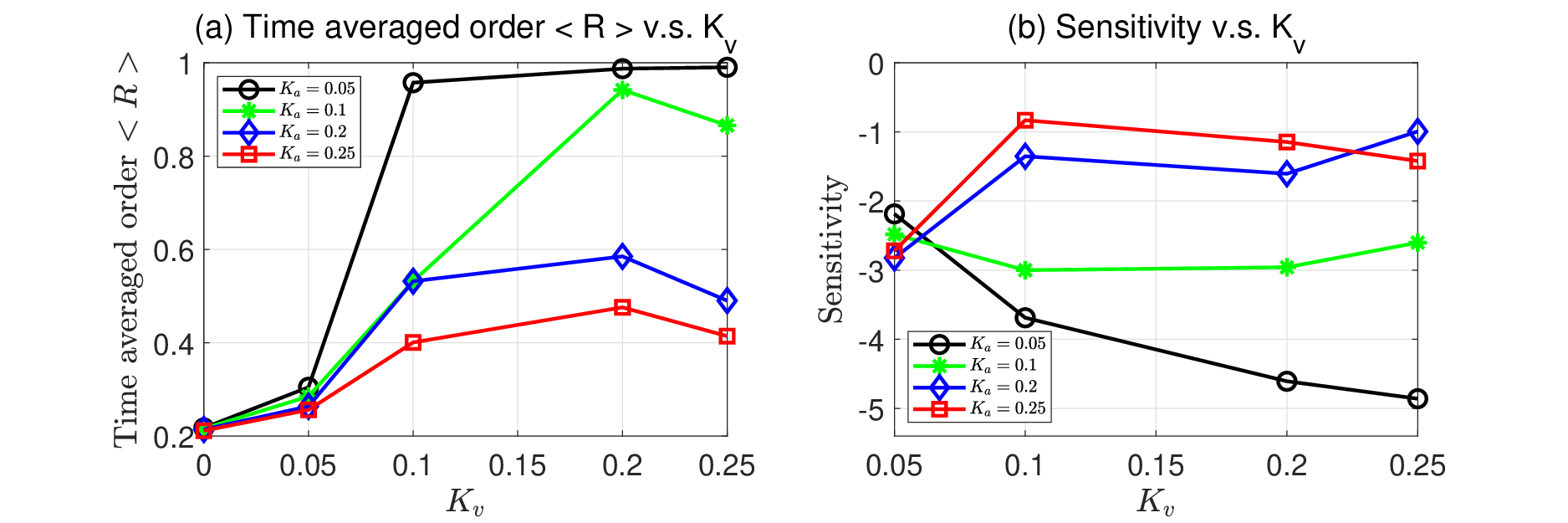}
            \caption{{\bf (a) The effects of $K_v$ on the time-averaged order and (b) its sensitivity measured by the relative mean-squared error (RMSE) for various values of $K_a$.} }
            \label{fig6}
            \end{figure}

\begin{figure}[!h]
            \centering
            \includegraphics[width=0.95\textwidth]{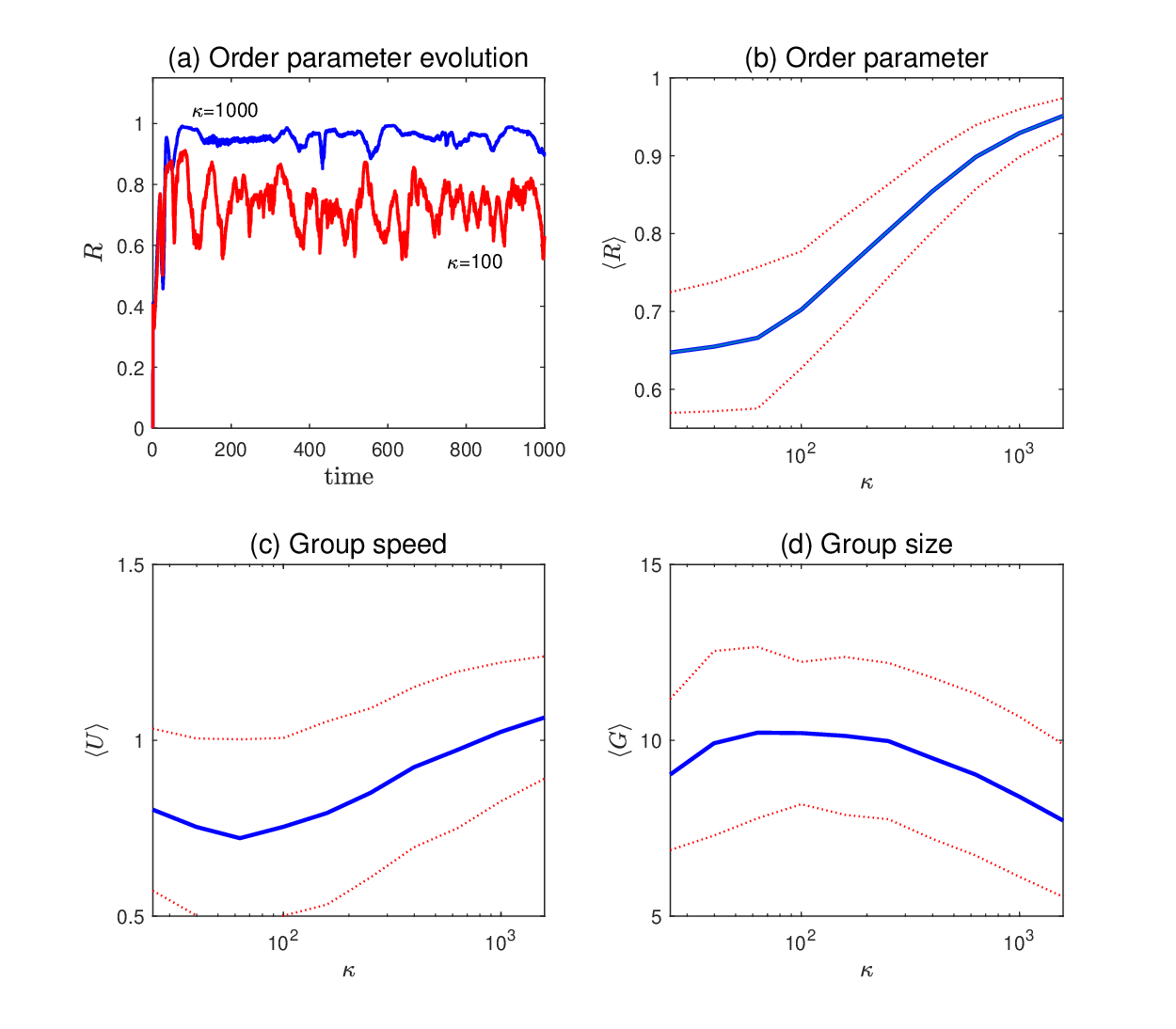}
            \caption{{\bf Macroscopic variables with respect to $\kappa$.}
            (a) Time evolution of the order parameter by varying the coefficient of the reaction time, $\kappa$. The time-averaged values of (b) the order parameter $\left<R\right>$, (c) the group speed $\left<U\right>$, and (d) the group size $\left<G\right>$, as a function of reaction time coefficient $\kappa$. $K_a=0.12$ is used. The red dotted lines in (b)-(d) indicate one standard deviations from the time-averaged data.}
            \label{fig7}
            \end{figure}

\begin{figure}[!h]
            \centering
            \includegraphics[width=0.95\textwidth]{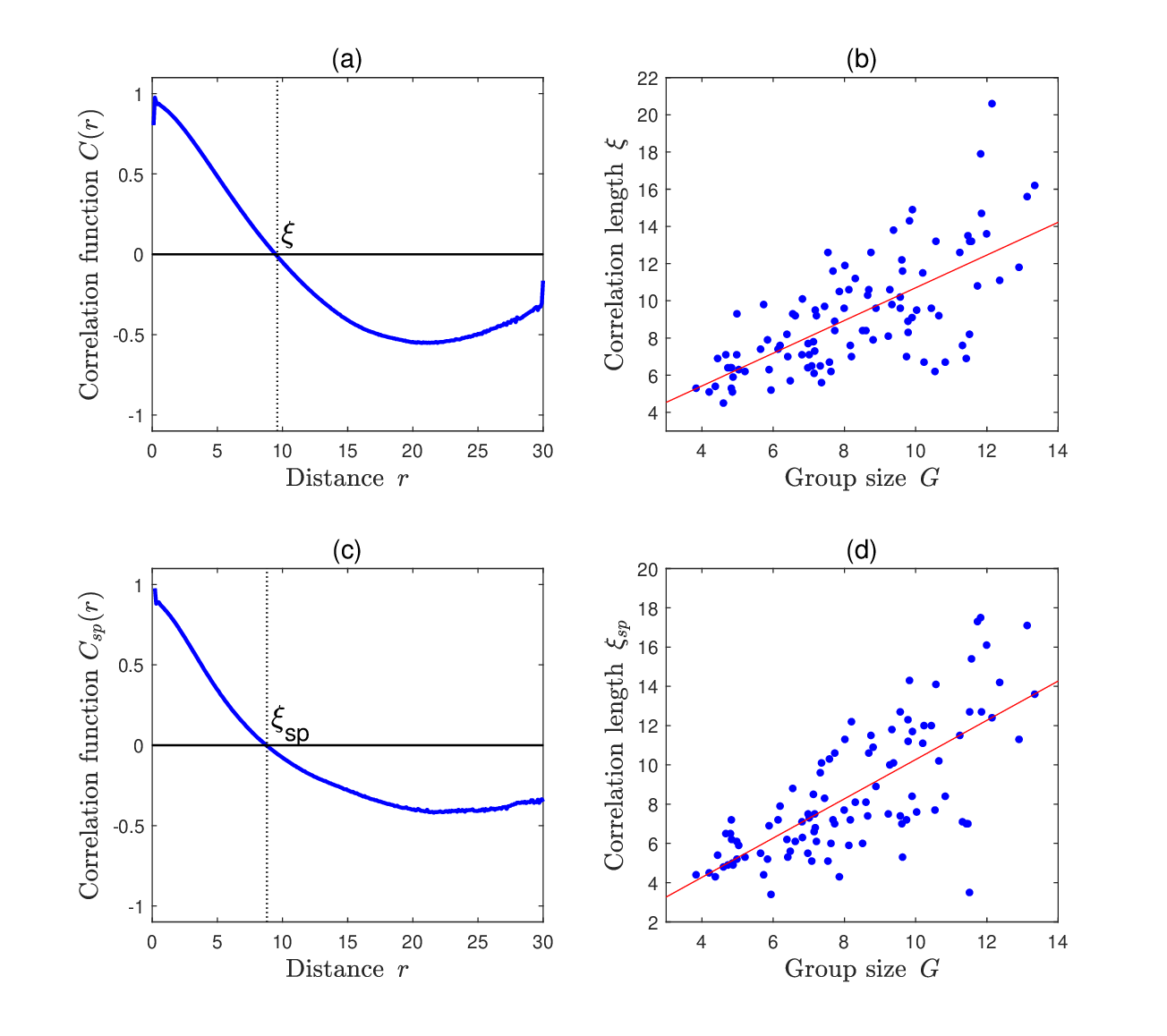}
            \caption{{\bf The correlation functions and the correlation lengths.}
            (a) and (b) for velocity, and (c) and (d) for speeds. The correlation lengths $\xi$ are denoted in (a) and (c), respectively, guided by the dotted lines. The data of 1000 points are sampled at equal intervals from one simulation run when $t_f=20000$ sec, which is larger enough than the early transient time and the parameter setting of the simulation is the same as in Fig~\ref{fig3}. The clear linear correlations are shown. (Person's correlation tests: $n=100$; for (b) $r=0.69$ and $p = 2\times10^{-15}$, and for (d) $r=0.72$ and $p < 10^{-15}$.}
            \label{fig8}
            \end{figure}

\textcolor{black}{In Fig. \ref{fig6}, we quantitatively investigate the relationship between time-averaged group orders $\left<R\right>$ and the sensitivity (the magnitudes of fluctuations) through the parameter $K_v$.
In the $K_v-$dominant regime where $K_a$ is as small as 0.05, the order increases to one as $K_v$ increases (Fig. \ref{fig6}(a)), while the magnitude of fluctuations monotonically decreases (Fig. \ref{fig6}(b)). A flock accordingly becomes non-responsive to the external perturbations in the $K_v-$dominant regime. Here, we measure the relative mean-squared error (RMSE) $\left<(R-\left<R\right>)^2\right>/\left<R\right>^2$ as the sensitivity indicator for a flock and present it in the log scale. 
Fig. \ref{fig6}(a) also depicts the $K_a-$dominant regime where $K_a$ is as large as 0.2 (blue diamonds) or 0.25 (red squares), and the order increases with $K_v$, but slowly only up to 0.6. On the other hand, the sensitivity increases due to the dominant role of $K_a$ shown in Fig. \ref{fig6}(b). This indicates that a flock becomes sensitive to external perturbations and therefore it gets poorly organized. 
Interestingly, when $K_a$ is intermediate around $K_a=0.1$ (green stars), its fluctuations are not completely suppressed while the order is maintained near one.}

%% Figure 4
%%[17]
%%(a)
\textcolor{black}{The optimal level of the sensitivity constant, $\kappa$, may differ from flock to flock of different species, giving different reaction times.} 
Figure \ref{fig7}(a) compares the time evolution of the group order $R$ at the two different values of $\kappa=100$ and 1000. 
As we previously mentioned, we choose two values of $\kappa$ such that both ones fall on an intermediate range, preventing excessive non-physical delays comparable to $t_f$. \textcolor{black}{According to \cite{ballerini2008interaction}, the scale of the delay in acceleration would be a couple of seconds for starlings in a large flock. In our study, we determine the value of $\kappa$ in an intermediate range from the corresponding average time delay $\tau$. In other words, $\kappa$ would be non-physical if $\tau$ is more than a couple of seconds. See the plot of the averaged $\tau$ vs $\kappa$ in the Supplementary Information.} We also exclude extremely short delays that simply lead to disordered states. The final simulation time is $t_f=10000$ sec, which is larger enough than the typical transient time (near $t=100$ sec) shown in the figure \ref{fig7}(a). 
    In the case with $\kappa=100$, the magnitude of the fluctuations is large and the group polarity is much lower than one ($\left<G\right>=0.65$), as shown (red) in Figure \ref{fig7}(a). 
        This is because rapid synchronization of the acceleration with others tends to increase uncertainty, making the bird's behavioral state deviate from the entire group's average dynamic state.
    At $\kappa=1000$, the fluctuation magnitudes are moderate and the group polarity is maintained at the value near one as shown (black) in Figure \ref{fig7}(a). 
        The elongated reaction time to others' acceleration decreases uncertainty that the bird's behavioral state mismatches with others. However, any ``urgent" fluctuations from neighbors that may occur faster than its reaction time are subject to be dampened away.
            These results indicate that the parameter $\kappa$ plays a role of the inverse of the temperature, similar to the stochastic noise $\eta$ in the classical Vicsek model \cite{vicsek1995novel}. 
            In other words, the deterministic rule of the individuals' adaptive reaction in our model can be accounted for by the stochastic noise term in Vicsek's model. 
            We emphasize that the birds' adaptive reactions enable the flock effectively to gain both polarity and responsiveness at the same time: this mechanism elicits sensitivity from the ordered steady state, by making the flock unstable at the highest polarity.

%%(b)-(d) setting
The time-averaged values of $\left<R\right>$, $\left<U\right>$, and $\left<G\right>$ are plotted as $\kappa$ increases in Figs~\ref{fig7}(b)--(d). (The symbols of a pair of angles denote the time average: $\left< \cdots \right>=\frac{1}{t_f} \int_0^{t_f} \cdots ~dt$). Their variances are also plotted in dotted lines (red).
%%R
    %1R
    The group polarity $\left< R \right>$ monotonically increases with $\kappa$, converging to one, as if the temperature is lowered. 
    \textcolor{black}{In fact, we can see that the elongated delay with the large $\kappa$ makes the collective dynamics more stable from the narrowed standard deviation in Fig. \ref{fig7}(b). This means that the corresponding macroscopic variables remain almost constant in time during the flight.}
    %%% 더 작업해야 함. 2
    We notice that there are two regimes of the macroscopic behaviors, which are divided by an extreme point of $\kappa$ in the plots of $\left< U \right>$ and $\left< G \right>$: 
    (i) the responsiveness-dominant regime where $\kappa$ is smaller than the extreme point and (ii) the ordering-dominant regime where $\kappa$ is larger than the point. 
    In regime (i), the degree of the responsiveness of birds makes differences in collective behaviors. 
    Whereas $\left< R \right>$ insignificantly changes in that regime of $\kappa$, $\left< U \right>$ decreases with $\kappa$ since birds become less responsive to others' behavioral changes and tend not to follow up others as much as in the case when $\kappa$ is smaller. 
    Due to the same tendency, $\left< G \right>$ decreases by the lowered birds' coherence. 
    %$\kappa$ is already too low and thus 
    %
    In regime (ii), the strength of birds to get ordered determines the characteristics of collective behaviors. 
    $\left< R \right>$ increases toward one as $\kappa$ increases. Consequently, as the group becomes better ordered, the faster the motions of its center ($\left<U\right>$ increases) and the smaller the group size ($\left<G\right>$ decreases) \cite{trenchard2012american}.

%% Figure Kv evolution
%% Four figures added

%% Figure 5
%%[18]
To statistically characterize the generated fluctuations, we examine the correlations of the fluctuations in velocity and speed, respectively, shown in Fig~\ref{fig8}. 
    The velocity fluctuations around the mean value is defined by 
    \begin{equation}
    \textbf{u}_i=\textbf{v}_i-\frac{1}{N}\sum_{k=1}^{N}\textbf{v}_k,
    \end{equation}
    where the sum of fluctuations around the group mean is zero as $\sum \textbf{u}_i=0$ by definition. 
    Similarly, the speed fluctuations with respect to the mean value is defined by 
    \begin{equation}
    {\phi}_i=||\textbf{v}_i||-\frac{1}{N}\sum_{k=1}^{N}||\textbf{v}_k||.
    \end{equation}
    %%[19]
    We next define correlation functions of fluctuations to measure how much two birds at a distance $r$ are correlated at a given time \cite{cavagna2010scale}. 
    The velocity correlation function is %
    \begin{equation}\label{corrvel}
    C(r)=\frac{1}{C_a}\frac{\sum_{ij}\textbf{u}_i \cdot \textbf{u}_j \delta(r-r_{ij})}{\sum_{ij}\delta(r-r_{ij})},
    \end{equation}
    where $\delta (r-r_{ij})$ is a Dirac $\delta$-function to select pairs of birds at mutual distance $r$, and $C_a$ is a normalization factor such that $C(r=0)=1$. 
    The correlation length $\xi$ is defined as a point that makes the correlation function zero, $C(r=\xi)=0$. It provides a good estimate of the average size of the correlated domain. 
    Similarly, we define the correlation function of fluctuations in speed to quantify the size of the correlated region:
    \begin{equation}\label{corrsp}
    C_{sp}(r)=\frac{1}{c_2}\frac{\sum_{ij}\phi_i \cdot \phi_j \delta(r-r_{ij})}{\sum_{ij}\delta(r-r_{ij})}
    \end{equation}
    where $c_2$ is a normalization factor such that $C_{sp}(r=0)=1$. 

%%
%%[20]
In Figs~\ref{fig8}(a), the correlation function of velocity, $C(r)$, is plotted with respect to the interindividual distance $r$. The corresponding correlation length of $\xi$ with $C(\xi)=0$ is denoted by the guidance of the vertical dotted line. 
To study its group size dependence, the correlation length $\xi$ as a function of the group size $G$ is presented in Figures ~\ref{fig8}(b). 
     Pearson's correlation test gives $n=100$ and $p=2 \times 10^{-15}$. The Pearson's correlation coefficient is $r=0.69$ close to $0.7$, where the two variables have a strong linear correlation for $r>0.7$.
%% Speed correlation
Similarly, the correlation function of speed $C_{sp}$ and the correlation length $\xi_{sp}$ with the group size $G$ are plotted in Figs~\ref{fig8}(c) and (d), respectively. 
    Pearson's correlation test gives $n=100$ and $p=10^{-15}$. The Pearson's correlation coefficient is $r=0.72 > 0.7$, implying strong linear correlation. 
    The fact that the correlation length is proportional to the group size implies that the correlation lengths are scale-free as experimentally reported \cite{cavagna2010scale}. 
    These results confirm that the model with the adaptive acceleration response generates scale invariance of the correlation lengths from the fluctuations in velocity and speed, respectively.  

\section*{Conclusion}
We have proposed an interindividual coordination mechanism for a collective response in a generalized Cucker-Smale model. 
We show that the adaptive reaction of a bird to acceleration that depends on the local polarity can create complex flocking patterns similar to those found in natural flocks. 
We found that at a given sensitivity level, the flock can maintain orientational order at a high level while responding to the perturbations of others. 
Furthermore, we confirm that the adaptive reaction mechanism generates the scale-free correlations in the fluctuations of both velocity and speed, as experimentally reported. 
These results indicate that the adaptive reaction keeps a flock in an optimal state so that they are sensitive to internal or external perturbations but not frozen in a steady state.

\begin{acknowledgments}
Narina Jung was supported by a KIAS Individual Grant (CG085701) at Korea Institute for Advanced Study and Basic Science Research Program through the National Research Foundation of Korea (NRF) funded by the Ministry of Education (Grant No. 2019R1A6A1A03033215). Byung Mook Weon was supported by Basic Science Research Program through the National Research Foundation of Korea (NRF) funded by the Ministry of Education (Grant No. 2019R1A6A1A03033215). Pilwon Kim was supported by the National Research Foundation of Korea (2017R1D1A1B04032921) and Ulsan National Institute of Science and Technology (1.200052.01). The funders had no role in study design, data collection and analysis, decision to publish, or preparation of the manuscript.
\end{acknowledgments}

\paragraph*{S1 Video.}
\label{S1_Video}
{\bf A flock of 2000 birds.} When $K_a=0.12$ and $\kappa=200$.

\paragraph*{S2 Video.}
\label{S2_Video}
{\bf A flock of 2000 birds.} When $K_a=0.12$ and $\kappa=800$.

\appendix

\section{Linear stability analysis of a two-bird system}

Equation (\ref{maineq1}) with the reaction time $\tau_{i}$ in response to the acceleration is one example of neutral delay differential equations, where a delay is considered in the terms with the highest order of derivative, i.e., acceleration in our case \cite{hale2013introduction,kuang1993delay,bray1966bifurcation}. In neutral delay differential equations, even minor delays can have significant effects on the stability of the systems \cite{kuang1993delay,kolmanovskii1986stability}.

Here we briefly present the standard stability analysis for the model in Eq  (\ref{maineq1}) without the fourth potential term when $N=2$. Given two birds, $i=1,2$, we assume that the two birds are flying with the same velocity $\textbf{v}_1=\textbf{v}_2=(v_x^*,v_y^*)^T$, and with the same reaction time $\tau_1=\tau_2=\tau$. Let $s$ be the distance between two birds, as $|\textbf{x}_2-\textbf{x}_1 |=s$. Note that, since the adjacent function $g(s)$ monotonically decreases, $g(s)$ grows as two birds are getting closer. Here we treat $g(s)$ as a parameter, assuming that the relative position of two birds $s$ is fixed in the analysis.

We can reformulate the velocity part of Eq (\ref{maineq1}) as
\begin{eqnarray}\label{redu}
\frac{d\textbf{v}_1}{dt}&=&\textbf{F}(\textbf{v}_1,\textbf{v}_2)+\textbf{G}\left(\tilde{\textbf{{a}}}_1,\tilde{\textbf{{a}}}_2\right) \\
\frac{d\textbf{v}_2}{dt}&=&\textbf{F}(\textbf{v}_2,\textbf{v}_1)+\textbf{G}\left(\tilde{\textbf{{a}}}_1,\tilde{\textbf{{a}}}_2\right) 
\end{eqnarray}
where $\textbf{F}(\textbf{u},\textbf{v})=K_vg(s)(\textbf{v}-\textbf{u})+\alpha(1-|\textbf{u}|^2)\textbf{u},$ 
$\textbf{G}(\textbf{a}_1,\textbf{a}_2)=K_a g(s) \textbf{a}_2$, 
$\tilde{\textbf{a}}_1(t)=\textbf{{a}}_1(t-\tau)$ 
and $\tilde{\textbf{a}}_2(t)=\textbf{a}_2(t-\tau)$. 
We set a solution of Eq (\ref{redu}) around the aligned formation $\textbf{v}_1=\textbf{v}_2=(v_x^*,v_y^*)^T$ as 
\begin{equation}
\textbf{y}(t)=\textbf{y}^*+\delta\textbf{y}(t)
\end{equation}
where $\textbf{y}=(v_{1x},v_{1y},v_{2x},v_{2y})^T$, $\textbf{y}^*=(v_{x}^*,v_{y}^*,v_{x}^*,v_{y}^*)^T$ and $\delta\textbf{y}$ is the infinitesimal displacements from the equilibrium solution.
Using the Taylor series expansion, the above Eq (\ref{redu}) can be linearized about the equilibrium solution as
\begin{equation}
\frac{d (\delta\textbf{y}) }{dt}=\textbf{J}\, \delta\textbf{y}+\textbf{J}_{\tau}\,\frac{d (\delta\tilde{\textbf{y}})}{dt}\label{lineqn}
\end{equation}
where $\delta\tilde{\textbf{y}}(t)=\delta\textbf{y}(t-\tau)$.
Here the Jacobian matrices $\textbf{J}$ and $\textbf{J}_\tau$ are
\begin{equation*}
\textbf{J}=
\begin{bmatrix}
-K_v g(s)-2 \alpha v_{x}^{*2} & -2 \alpha v^*_{x}v^*_{y} & K_v  g(s) & 0 \\
-2\alpha v^*_{x}v^*_{y} & -K_v  g(s)-2 \alpha v_{y}^{*2} & 0 & K_v  g(s) \\
K_v  g(s) & 0 & -K_v  g(s)-2\alpha v_{x}^{*2} & -2\alpha v^*_{x}v^*_{y} \\
0 & K_v  g(s) & -2\alpha v^*_{x}v^*_{y} & -K_v  g(s)-2\alpha v_{y}^{*2} 
\end{bmatrix}
\end{equation*}
and
\begin{equation*}
\textbf{J}_\tau=
\begin{bmatrix}
0 & 0 & K_a  g(s) & 0 \\
0 & 0 & 0 & K_a  g(s) \\
 K_a  g(s) & 0 & 0 & 0 \\
0 &  K_a  g(s) & 0 & 0 
\end{bmatrix}.
\end{equation*} 

We seek exponentially growing solutions of (\ref{lineqn}) of the form
\begin{equation}\label{linsol}
\delta\textbf{y}(t)=e^{\lambda t} \textbf{w}, \,\textbf{w}\ne 0
\end{equation}
where $\lambda$ is complex and $\textbf{w}$ is a vector whose components are complex. Putting Eq (\ref{linsol}) to Eq (\ref{lineqn}) gives a characteristic equation with respect to $\lambda$ as
\begin{equation}
\label{chareqn}
\begin{aligned}
0=&\text{det}(\textbf{J}+\lambda e^{\lambda t} \textbf{J}_\tau -\lambda \textbf{I})\\
=&\lambda e^{-4\lambda\tau}(e^{\lambda\tau}-K_a g(s))(\lambda K_a g(s)+\lambda e^{\lambda\tau}+2K_v g(s) e^{\lambda\tau}+2\alpha e^{\lambda\tau})\\
&(-\lambda K_a g(s)+\lambda e^{\lambda\tau}+2\alpha e^{\lambda\tau})(\lambda K_a g(s)+\lambda e^{\lambda\tau}+2K_v g(s) e^{\lambda\tau}+2\alpha  e^{\lambda\tau}) 
\end{aligned}
\end{equation}
where $\textbf{I}$ is a $4\times 4$ identity matrix. The five factored equations for the eigenvalues in (\ref{chareqn}) are
\begin{eqnarray}
0&=&\lambda e^{-4\lambda\tau}\label{e1}\\
0&=&e^{\lambda\tau}-K_a g(s)\label{e2}\\
0&=&\lambda K_a g(s)+\lambda e^{\lambda\tau}+2K_v g(s) e^{\lambda\tau}+2\alpha e^{\lambda\tau}\label{e3}\\
0&=&-\lambda K_a g(s)+\lambda e^{\lambda\tau}+2\alpha e^{\lambda\tau}\label{e4}\\
0&=&\lambda K_a g(s)+\lambda e^{\lambda\tau}+2K_v g(s) e^{\lambda\tau}+2\alpha  e^{\lambda\tau}\label{e5} 
\end{eqnarray}  

Let $\lambda^{\text{Re}}_{\text{max}}$ denote the largest value of the real part of eigenvalues of the linearlized system. For the system to be stable, $\lambda^{\text{Re}}_{\text{max}}$ should be nonpositive. One can confirm that there is no positive solution of the real part eigenvalue from the below three equations (\ref{e3}),(\ref{e4}) and (\ref{e5}). From the first two equations (\ref{e1}) and (\ref{e2}), we have
\begin{equation}\label{lambda_comp}
\lambda^{\text{Re}}_{\text{max}}=
\begin{cases}
0, & \text{if } K_a g(s) \leq 1, \\
\frac{1}{\tau}\log(K_a g(s)), & \text{if } K_a g(s) > 1.
\end{cases}
\end{equation}

\begin{figure}[!h]
            \centering
            \includegraphics[width=0.95\textwidth]{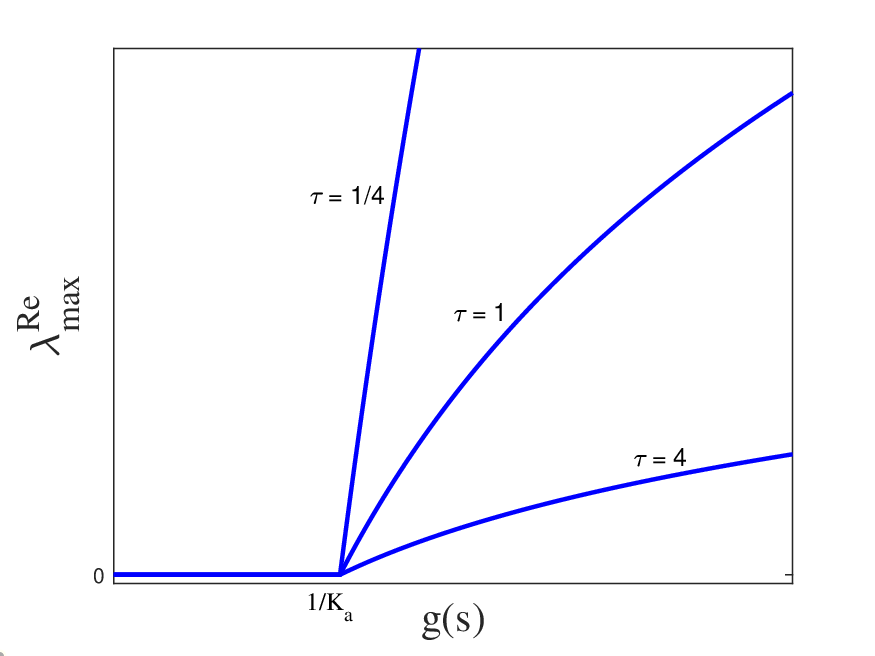}
            \caption{Maximal eigenvalue according to the communication rate $g(s)$, where $s$ is a distance between two birds as $s=|\textbf{x}_2-\textbf{x}_1|$.  }
            \label{fig9}
            \end{figure}

Figure~\ref{fig9} plots the maximum eigenvalue with respect to the communication rate $g(s)$. The value of $\lambda^{\text{Re}}_{\text{max}}$ bifurcates from a neutral state to an unstable one at a critical value $K_a g(s) = 1$, and the system is unstable when $K_a g(s) > 1$. Since the communication rate $g(s)$ monotonically decreases with $s$, the infinitesimal perturbation of the two birds away from the aligned position at equilibrium becomes unstable when $s < s_c$ where $K_a g(s_{\text{c}}) = 1$. 

%% Include only the SI item label in the paragraph heading. Use the \nameref{label} command to cite SI items in the text.

%
%\paragraph*{S1 Appendix.}
%\label{S1_Appendix}
%{\bf Lorem ipsum.} Maecenas convallis mauris sit amet sem ultrices gravida. Etiam eget sapien nibh. Sed ac ipsum eget enim egestas ullamcorper nec euismod ligula. Curabitur fringilla pulvinar lectus consectetur pellentesque.
%
%\paragraph*{S1 Table.}
%\label{S1_Table}
%{\bf Lorem ipsum.} Maecenas convallis mauris sit amet sem ultrices gravida. Etiam eget sapien nibh. Sed ac ipsum eget enim egestas ullamcorper nec euismod ligula. Curabitur fringilla pulvinar lectus consectetur pellentesque.

\textcolor{black}{\section{The comparison with the adaptive velocity and the constant acceleration equations}}

\begin{figure}[!h]
            \centering
            \includegraphics[width=0.95\textwidth]{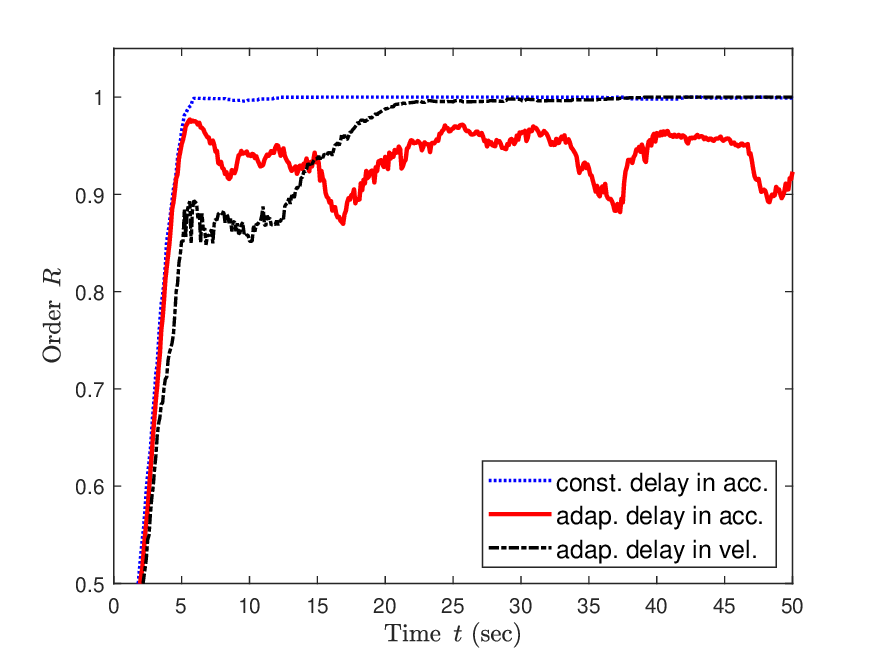}
            \caption{The comparison with the adaptive velocity and the constant acceleration equations  }
            \label{fig10}
            \end{figure}

%Figure Comparison
\textcolor{black}{To show the role of the adaptive delay in our model, we include the comparison of the following three cases in Fig. \ref{fig10}: (1) a constant delay in acceleration (when $\tau_i=\tau_0$ for all $i$ and parameters are $K_a=0.1$, $K_v=0.2$, $\kappa=800$), (2) an adaptive delay in acceleration (parameters are $K_a=0.1$, $K_v=0.2$, $\kappa=800$), and (3) an adaptive delay in velocity (parameters are $K_a=0$, $K_v=0.2$, $\kappa=800$). Note that the case (1) is the model similar to the work by Szabo et al. \cite{szabo2009transitions}. 
Since the instantaneous reaction near $R=1$ induces instability, the adaptive acceleration (solid red) prevents the system from converging into a formation that is perfectly aligned with $R=1$. For the cases of (1) (dotted blue) and (3) (dash-dot black), alignment states converge to the well-ordered configuration and remains in that steady state for the last of the evolution time. This comparison indicates that the order-dependent delay in acceleration is the essential factor in generating a rich dynamics of a flock, providing the response sensitivity of the flock from the external perturbations.} 

\bibliography{ART_ref.bib}% Produces the bibliography via BibTeX.

\end{document}